\documentclass[prl, twocolumn]{revtex4}
\usepackage{amsmath}
\usepackage{graphicx}
\usepackage{xcolor}
\def\lsim{\mathrel{\raise3pt\hbox to 8pt{\raise -6pt\hbox{$\sim$}\hss{$<$}}}}

\begin{document}
\bibliographystyle{apsrev}

\title{Analysis of the Daya Bay Reactor Antineutrino Flux Changes with Fuel Burnup }

\author{A.C. Hayes, Gerard Jungman} 
\affiliation{Los Alamos National Laboratory, Los Alamos, NM, 87545, USA}
\author{ E.A. McCutchan, A.A. Sonzogni}
\affiliation{National Nuclear Data Center, Brookhaven National Laboratory, Bldg. 817, Upton, NY, 11973-500, USA}
\author{G.T. Garvey}
\affiliation{University of Washington, Seattle, WA, 98195, USA}
\author{X.B. Wang} 
\affiliation{School of Science, Huzhou University, Huzhou 313000, China}
\begin{abstract}
We investigate the recent Daya Bay results on the changes in the antineutrino flux and spectrum with the burnup of the reactor fuel.
We find that the discrepancy between current model predictions and the Daya Bay results can be traced to the
original measured $^{235}$U/$^{239}$Pu ratio of the fission beta spectra that were used as a base for the expected antineutrino fluxes. 
An analysis of the antineutrino spectra that is based on a summation over all fission fragment beta-decays, using nuclear database input, explains all of the features seen in the Daya Bay evolution data. However,
this summation method still predicts an anomaly.
Thus, we conclude that there is currently not enough information to use the antineutrino flux changes
 to rule out the possible existence of
sterile neutrinos.
\end{abstract}
\maketitle

Recent results from  the Daya Bay (DB) reactor neutrino experiment \cite{DayaBay-new} show significant change in 
the emitted antineutrino flux with the evolution of the reactor 
fuel. Over the course of 
 1230 days, the fuel evolved such that the fraction of fissions from $^{239}$Pu increased from 25\% to 35\%, while those from 
$^{235}$U decreased from 63\% to 51\%. Over the same period, the fraction from $^{238}$U remained approximately constant at 7.6\%, while the 
$^{241}$Pu fraction increased from 4\% to 8\%.
The dependence of antineutrino flux on the fuel evolution was measured \cite{DayaBay-new} by the change in the yield from 
the inverse beta decay (IBD) reaction $\nu+p\rightarrow e^++n$  with the variation in the $^{239}$Pu fission fraction, $F_{239}$. 
 The IBD yield, which is an integral over energy of the product of the IBD cross section and the antineutrino flux per fission,  was fitted with a linear dependence on $F_{239}$ as \cite{DayaBay-new},
\begin{equation}
\sigma_f(F_{239})=\overline{\sigma}_f + \frac{d\sigma_f}{dF_{239}}\left(F_{239}-\overline{F}_{239}\right),
\end{equation}
where $\overline{\sigma}_f$ is the average IBD yield, $\overline{F}_{239}$ is the average $^{239}$Pu fission fraction, and 
$\frac{d\sigma_f}{dF_{239}}$ is the change of the IBD yield per unit $^{239}$Pu fission fraction. 
The values reported by Daya Bay are: $\overline{\sigma}_f=(5.9\pm0.13)\times 10^{-43}$ cm$^2$/fission, 
 $\frac{d\sigma_f}{dF_{239}}=(-1.86 \pm 0.18) \times 10^{-43}$ cm$^2$/fission, and $\overline{F}_i=(0.571,0.076,0.299,0.054)$
 for i=($^{235}$U, $^{238}$U, $^{239}$Pu, $^{241}$Pu).

These DB results confirm the ``reactor neutrino anomaly'' \cite{anomaly, daya-bay}, in that the measured value of 
 $\overline{\sigma}_f$
is about 5.1\% below that predicted by the model spectra of Huber and Mueller (H-M) \cite{huber, mueller}. 
However, the new DB results question the origin of this anomaly because the
magnitude of the anomaly varies with the fuel evolution. The variation in the size of the anomaly with the fuel evolution
results from the fact that the H-M value for $ \frac{d\sigma_f}{dF_{239}}=(-2.46\pm0.06) \times 10^{-43}$ cm$^2$/fission differs
 from DB's measured  value by 3.1 $\sigma$. 
The H-M ratio ${\overline{\sigma}_f}/\frac{d\sigma_f}{dF_{239}}$ does not agree with experiment and
 is incompatible with the IBD deficit being the same for all four actinides by 2.6 $\sigma$. 
DB's experimentally deduced IBD yields for $^{235}$U and $^{239}$Pu are $\sigma_{235}=(6.17\pm0.17) \time 10^{-43}$ cm$^2$/fission
 and $\sigma_{239}=(4.27\pm0.26) \times 10^{-43}$ cm$^2$/fission, 
respectively, corresponding to a 
$\sigma_{235}/\sigma_{239}$ ratio of 1.445$\pm 0.097$. By comparison, the Huber model ratio is 1.534$\pm$ 0.05.
The DB analysis \cite{DayaBay-new} suggests that the
anomaly arises almost entirely from $^{235}$U, and that the Huber prediction \cite{huber} for IBD yield for $^{235}$U, $\sigma_{235}$, is 7.8\% larger than that deduced by DB, 
while the model IBD yield for $^{239}$Pu, $\sigma_{239}$, is in reasonable agreement with experiment.  

The purpose of the present work is to point out that (1) the Huber prediction for $\sigma_{235}/\sigma_{239}$ is strongly constrained by the
 original measured aggregate beta spectra of Schreckenbach {\it et al.} \cite{schreck} that Huber converted to antineutrino spectra, 
and (2) a nuclear database analysis, involving a summation over all beta-decay transitions
that make up the aggregate antineutrino spectra, 
 provides a reasonable description of all of the evolution data, but still predicts an anomaly.
Thus, it is difficult to draw a conclusion about the existence of sterile neutrinos from evolution data alone.

The experimental aggregate beta spectra were obtained in the 1980's \cite{schreck} at the Institute Laue-Langevin (ILL).
To investigate the origin of the Huber $\sigma_{235}/\sigma_{239}$ ratio, we refitted the ILL beta decay spectra, varying many
 of the
assumptions that go into such a  fit. 
The spectra were fitted assuming different combinations of allowed and first forbidden beta transitions, 
ranging from all allowed to 40\% first forbidden. 
The procedure and parameterization that we employed is described in \cite{hayes}. 
Only 25 or so transitions are required to fit the integral beta spectra. Thus, 
in order to calculate the the Fermi function and its finite size correction, a choice must be made to assign a $Z_{\rm eff}$ and $A_{\rm eff}$
 to these effective transitions. 
These choices of $Z_{\rm eff}$ and $A_{\rm eff}$ and the related endpoint energies introduce uncertainty into the fit, with a corresponding
uncertainty in the antineutrino spectra. 
Thus, in fitting the spectra the prescriptions for $Z_{\rm eff}$ and $A_{\rm eff}$ were also varied. 
The relative importance of the different approximations used in deriving expected antineutrino spectra is summarized in \cite{hayes-vogel}.    

Varying all of the assumptions in fitting the aggregate fission beta spectra for $^{235}$U and $^{239}$Pu led to variations in 
the corresponding antineutrino spectra that differed at the few percent level. 
However, in all cases the ratio of the antineutrino spectra and IBD yield ratio varied only slightly, with $\sigma_{235}/\sigma_{239}$ remaining
 close to 1.53, Fig. 1 and Table 1. 
In this figure  and table we show results for four sets of assumptions: 
(1) all transitions are allowed and Huber's quadratic prescription for $Z_{\rm eff}$,
(2) all transitions are allowed and $Z_{\rm eff}=\Sigma Y_{c_i} Z_i/\Sigma Y_{c_i}$, 
(3) transitions can be either  
allowed or forbidden and  $Z_{\rm eff}=\Sigma Y_{c_i}Z_i/\Sigma Y_{c_i}$, and (4) transitions can be either allowed or 
forbidden and $Z_{\rm eff}=\sqrt{\Sigma Y_{c_i}Z_i^2 / \Sigma Y_{c_i}}$. Here $Y_{c_i}$ are the cumulative fission yields for the
fission fragments $(Z_i, A_i)$.
We find that, for all sets of assumptions that we checked, the
 fits to the Schreckenbach beta spectra result in an IBD yield ratio with 
 $\sigma_{235}/\sigma_{239}$ that is about 6\% higher than the DB result.
\begin{table}
\caption{The individual IBD cross sections $\sigma_{235}$ and $\sigma_{239}$ change by a few percent
when the assumptions in fitting the ILL  aggregate beta spectra are changed.
But the ratio $\sigma_{235}/\sigma_{239}$ always remains close to 1.53}
\begin{tabular}{l|c|c|c|c}
\hline
&all allowed& all allowed&allow.+forbid.&allow.+forbid.\\
&$Z_{\rm eff}^{\rm Huber}$&$Z_{\rm  eff}$&$Z_{\rm eff}$&$(Z_{\rm eff}^2)^{1/2}$\\
\hline
$^{235}$U&6.69&6.58&6.47&6.48\\
$^{239}$Pu&4.36&4.3&4.22&4.23\\
ratio&1.534&1.530&1.533&1.532\\\hline
\end{tabular}
\end{table}

\begin{figure}
\includegraphics[width=7.5cm] {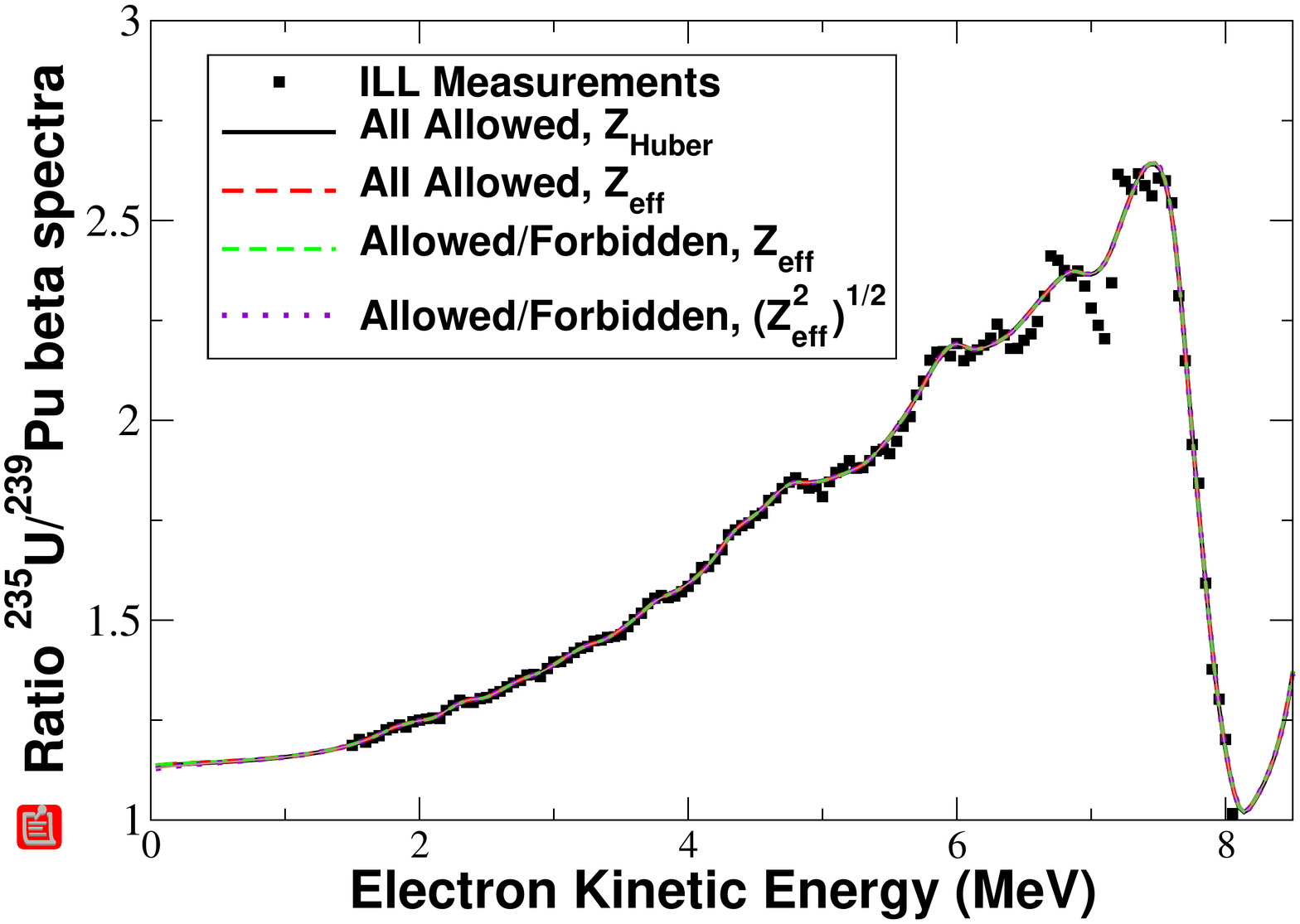}\\
\includegraphics[width=7.5cm] {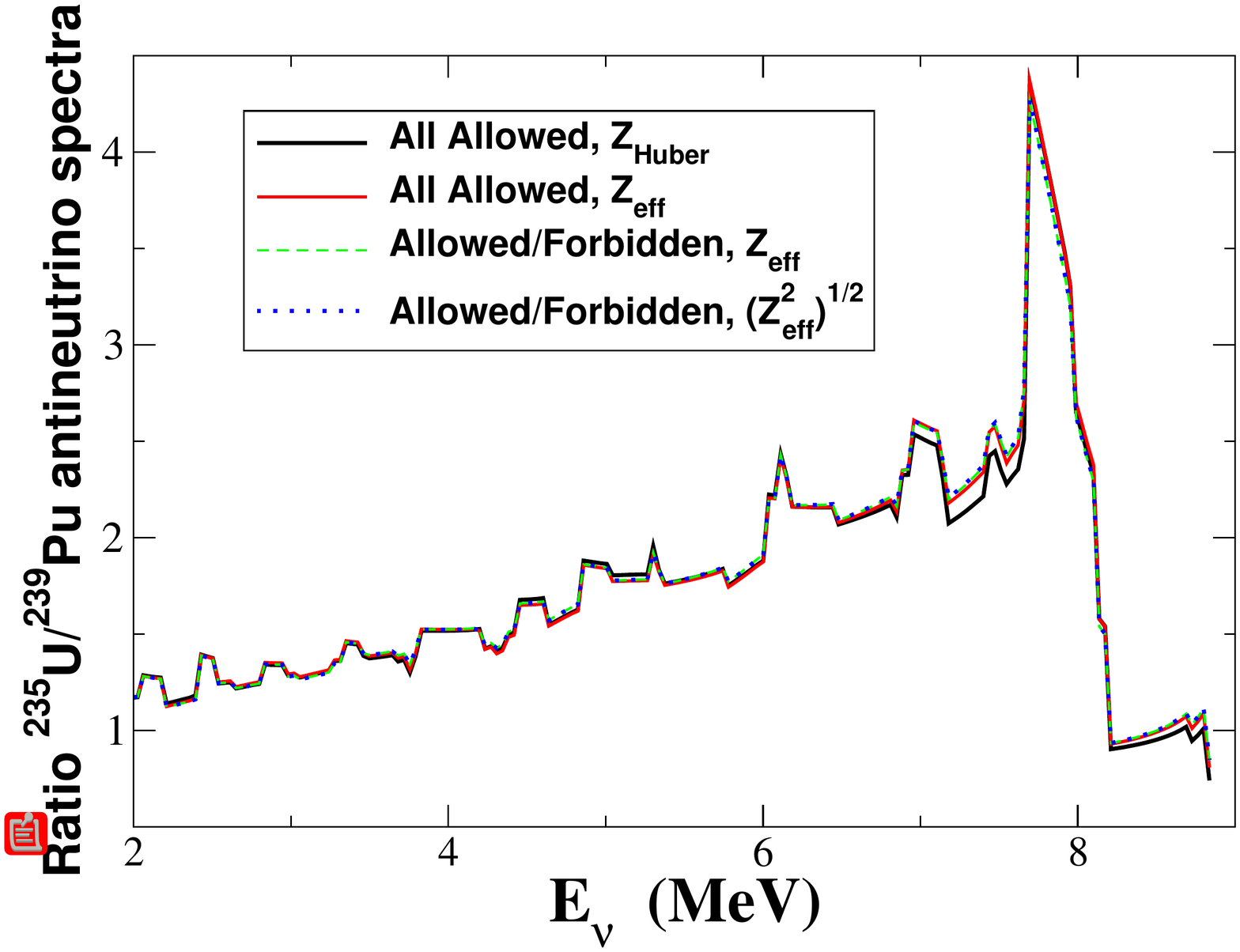}
\caption{(top)The ratio of the $^{235}$U and $^{239}$Pu beta spectra. The data are from  
\cite{schreck},
and the curves are the ratios obtained by fitting the individual $^{235}$U and $^{239}$Pu data, using different assumptions.
 The different assumptions are explained in the text.
 Excellent fits were obtained in all cases. 
(bottom)The ratio of the antineutrino spectra resulting from the fits. 
We note that the jagged structures largely reflect the fact that the fits  only require about 25 endpoints; these effects are normally smoothed in published expected spectra.}

\label{fit}
\end{figure}

An alternate procedure for investigating the $\sigma_{235}/\sigma_{239}$ ratio is to employ
the so-called summation method using the nuclear database libraries for the cumulative fission yields and beta decay spectra. 
In this work we have used the JEFF-3.1 cumulative fission yields \cite{JEFF}
 in combination with a preliminary version of the ENDF/B-VIII.0 decay data sub-library \cite{endf-8}
 as described in Ref. \cite{bnl}.  
ENDF/B fission yields were not used due to the compatibility issues discussed in Ref. \cite{BNL-bump}.   
For most of the energy interval, 2-7 MeV, these summation calculations predict a smaller $^{235}$U/$^{239}$Pu 
beta spectra ratio, see Fig. 2, leading to an IBD antineutrino yield ratio equal to 1.46.   
However, it is difficult to draw any conclusions from this fact because about 4\% 
of the predicted $^{235}$U electron spectra and 7\% of the $^{239}$Pu predicted electron spectra originate from nuclei whose decays are quite uncertain. 
In such cases the theoretical spectra of Kawano {\it et al.} \cite{kawano} were used.   
In addition,  the uncertainty on the database summation spectra was not estimated because
 correlation matrices for fission yields are not available.
The summation method prediction for $\frac{d\sigma_f}{dF_9}$, which also involves $^{238}$U and $^{241}$Pu, is in closer agreement
 with the Daya Bay result than the H-M model, Table 2 and Fig. 3.
However, the DB and summation results differ in detail.
In particular, the summation predictions for the IBD cross section for $^{235}$U, $^{239}$Pu and $^{241}$Pu are all about 5\% higher than the Daya Bay values.
Thus, all three actinides contribute approximately equally to the summation anomaly. In the case of $^{238}$U,
the uncertainty in the antineutrino spectrum is larger because $^{238}$U
 involves fast (as opposed to thermal)  fission yields. In addition, $F_{238}$ does not change
significantly with the fuel evolution.   
\begin{figure}[h]
\includegraphics [width=7.5 cm]{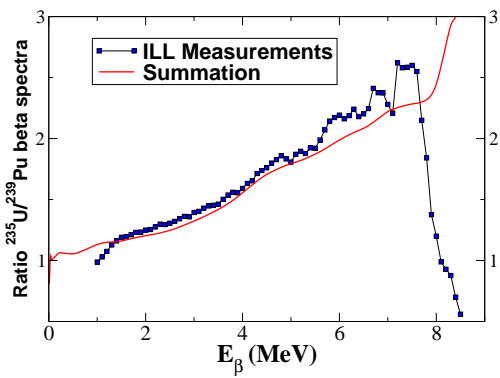}
\caption{The ratio of the $^{235}$U to $^{239}$Pu aggregate beta spectra as a function of the kinetic energy of the electron, for the
Schreckenbach {\it et al.} \cite{schreck} measurement (squares), and the summation method (curve).}
\label{ratio}
\end{figure} 
\begin{table}
\begin{tabular}{c|c|c|c}
\hline
&DB$^a$&Summation&H-M$^b$ \\\hline
$\overline{\sigma}_f$(10$^{-43}$cm$^2$)&5.9$\pm$ 0.13&6.11&6.22$\pm$0.14\\
$\frac{d\sigma_f}{dF_{239}}$(10$^{-43}$cm$^2$)&-1.86$\pm$ 0.18&-2.05&-2.46$\pm$0.06\\
&  & &\\\hline
$\sigma_5$ (10$^{-43}$cm$^2$)&6.17$\pm$ 0.17&6.49&6.69$\pm$0.15\\
$\sigma_9$ (10$^{-43}$cm$^2$)&4.27$\pm$ 0.26&4.49&4.36$\pm$0.11\\
$\sigma_8$ (10$^{-43}$cm$^2$)&10.1$\pm$1.0&10.2&10.1$\pm$1.0\\
$\sigma_4$ (10$^{-43}$cm$^2$)&6.04$\pm$0.6&6.4&6.04$\pm$0.6\\\hline
$\sigma_5/\sigma_9$&1.445$\pm$0.097&1.445&1.53$\pm$ 0.05\\
\hline
\end{tabular}
\caption{The IBD average yields, the variation with the $^{239}$Pu content of the fuel, and the contributions from individual actinides.  
$^a$The DB values for $\sigma_8$ and  $\sigma_4$ were assumed. 
$^b$ The uncertainties quoted for the H-M model are those used by the DB collaboration.
 A more direct comparison between the summation
 predictions and  experimental IBD yield data is shown in Fig.3-5.}
\end{table}
\begin{figure} [h]
\includegraphics[width=7.5 cm]{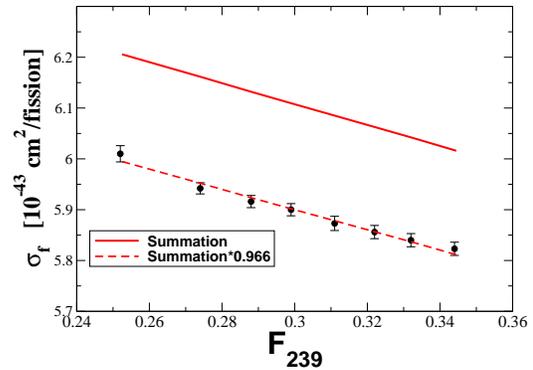}
\caption{The IBD yield per fission as a function of the fraction of fissions from $^{239}$Pu. The data are from Daya Bay \cite{DayaBay-new}, while the straight (dashed) curves are the absolute (renormalized) predictions from
the summation calculations.
The slope of the summation predictions for the change in the the IBD yield with F$_{239}$ is in agreement with experiment, but
the absolute value of the predicted IBD yield is 3.5\% high.}
\label{sigma}
\end{figure}

The  Daya Bay collaboration also observed a change in the shape of antineutrino spectrum over the course of the reactor fuel evolution.
This is defined  as $\frac{1}{S_j}\frac{dS_j}{dF_{239}}$, where $j$ denotes four prompt energy intervals $E^j_p$, 
(0.7-2 MeV, 2-4 MeV, 4-6 MeV, and 6-8 MeV), with $E_p=E_\nu+0.8$ MeV. 
$S_j$ is the corresponding partial contribution to the IBD yield in the energy range $E^j_p$:
\begin{equation}
S_j(F_{239})=\overline{S}_j+\frac{dS_j}{dF_{239}}(F_{239}-\overline{F}_{239})\;.
\end{equation} 
The summation predictions, along with the DB measurements are shown in Fig. 4, where good agreement is seen. 
\begin{figure}[h]
\includegraphics[width=9.0 cm]{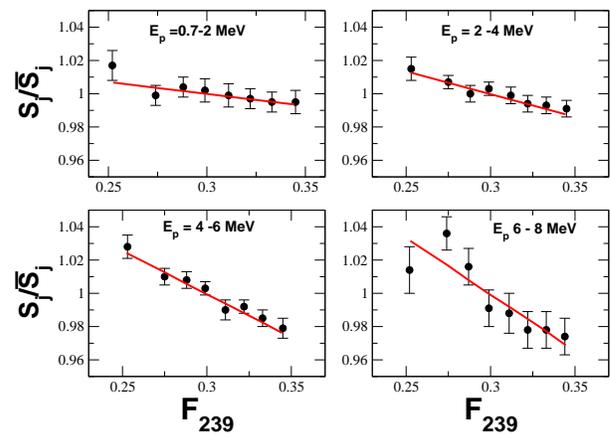}
\caption{The variation of the IBD yield in four prompt energy ranges. The data are from \cite{DayaBay-new}, while the straight lines are
the predictions of the summation method. 
The summation predictions for $\overline{S}^{-1}dS_j/dF_{239}$ are 
(-0.143, -0.273, -0.521, -0.678) for j=1-4, 
to be compared with the experimental values of (-0.16$\pm$0.07, -0.23$\pm$0.04, -0.49$\pm$0.05, -0.69$\pm$0.12).}
\label{S-values}
\end{figure}
A comparison to the change in the IBD spectrum with $F_{239}$ for six prompt energy ranges is shown in Fig.5.
In this figure we show both the summation
 predictions and one of our conversions of the ILL data, using  assumption (2)
 of Fig. 1.
The current fit to ILL leads to a change in the IBD spectrum that is very similar to the Huber model, while the summation
 predictions are closer to experiment.
\begin{figure}[h]
\includegraphics[width=7.5cm]{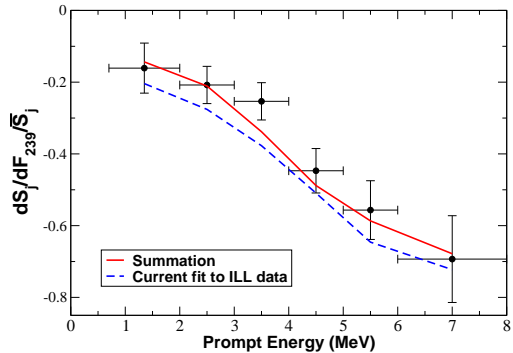}
\caption{The variation of the IBD yield for different prompt energy ranges. The data are from \cite{DayaBay-new},  the solid line is the
 prediction of the summation method, while the dashed line is obtained from converting the ILL data to antineutrino spectra and using the database for $^{238}$U.
}
\label{Six-values}
\end{figure}

The Daya Bay collaboration concluded that the expected Huber model $^{235}$U spectrum is too high in magnitude, while that for $^{239}$Pu
is consistent with the DB data.
This raises the question whether the measured changes in IBD yield and spectrum are consistent with a sterile neutrino explanation 
of the reactor neutrino anomaly.
The present analysis suggests that there is currently insufficient evidence to draw any conclusions on this issue.
As we have shown, an analysis based on the summation method
 explains all of the features seen in the evolution data,
but it predicts an average IBD yield that is 3.5\% higher than observed.
All actinides except $^{238}$U contribute approximately equally to the summation anomaly. But we note that 
$^{238}$U does not evolve with the rest of the fuel, and
its summation antineutrino spectrum is at least 10\% uncertain.
Resolving the issue of the existence of sterile neutrinos requires new very short baseline neutrino experiments.
A re-measurement of the aggregate fission beta spectra of $^{235}$U and $^{239}$Pu would also be very valuable in determining whether there
is a problem with the $\sigma_{235}/\sigma_{239}$ ratio.
\acknowledgments
{The research at Los Alamos National Laboratory was sponsored by the U.S. Department of Energy FIRE Topical Collaboration.
The research at Brookhaven National Laboratory was sponsored by the Office of Nuclear Physics, Office of Science of the U.S. Department of Energy under Contract No. DE-AC02-98CH10886.
X.B. Wang  was sponsored by the National Natural Science Foundation of China under Grants 
No. 11505056 and No. 11605054 and China Scholarship Council (201508330016).} 


\begin{thebibliography}{99}
\bibitem{endf-8}The ENDF/B-VIII.0 library, which is soon to be released, includes all TAGS beta decay data published \cite{TAGS} since the release of
ENDF/B-VII.1 \cite{endf-7}.
\end{thebibliography}
\end{document}